\documentclass[12pt,preprint]{aastex}

\shorttitle{Dusty Sources at the GC}
\shortauthors{Viehmann et al.}  

\begin{document}

\title{Dusty Sources at the Galactic Center:\\
       The $N$- and $Q$-band view with VISIR}

\author{
T.~Viehmann\altaffilmark{1} \and A.~Eckart\altaffilmark{1} \and
R.~Sch\"odel\altaffilmark{1} \and J.-U.~Pott\altaffilmark{1,2} \and 
J.~Moultaka\altaffilmark{1} 
}
\altaffiltext{1}{I.~Physikalisches Institut, Universit\"at zu K\"oln,
Z\"ulpicher~Str.~77, D-50937 K\"oln, Germany \email{viehmann@ph1.uni-koeln.de}}
\altaffiltext{2}{ESO, Karl-Schwarzschild-Str.~2, D-85748 Garching, Germany}

\begin{abstract}
We present mid-infrared $N$- and $Q$-band photometry of the Galactic Center from
images obtained with the mid-infrared camera VISIR at the ESO VLT in May 2004.
The high resolution and sensitivity possible with VISIR enables us to
investigate a total of over 60 point-like sources, an unprecedented number for
the Galactic Center at these wavelengths.
Combining these data with previous results at shorter wavelengths
\citep{viehmann2005} enables us to construct SEDs covering the $H$- to $Q$-band
regions of the spectrum, i.e. 1.6 to 19.5~$\mu$m.
We find that the SEDs of certain types of Galactic Center sources show
characteristic features.
We can clearly distinguish between luminous Northern Arm bow-shock sources,
lower luminosity bow-shock sources, hot stars, and cool stars.
This characterization may help clarify the status of presently unclassified
sources.
\end{abstract}




\section{Introduction}
\label{INTRO}

The Galactic Center is known as a bright source of near- and mid-infrared
radiation since the late 1960s \citep{becklin1968, becklin1969, low1969}.
The main source of near-infrared radiation is photospheric emission from a
dense stellar cluster, i.e.\ a crowded field of point sources,
while almost all of the mid-infrared radiation originates from extended gas and
dust features as well as dust emission from the circumstellar regions of a dozen
individual sources interacting with the more extended Galactic Center
interstellar medium.

The nature of the brightest mid-infrared sources is more or less clear --
IRS~1W, 2, 5, 10W, and 21 are bow-shock sources, caused by bright stars with
strong winds plowing through the ambient gas and dust of the Northern Arm
\citep{tanner2002, tanner2003, tanner2005, rigaut2003, eckart2004, geballe2004}.
Of the fainter sources, IRS~7 is an M2 supergiant, however, most of the others
remain enigmatic.

Combined photometry at near- and mid-infrared wavelengths is an ideal tool
to investigate the nature of the bright mid-infrared sources, since it gives
information on both the mid-infrared sources themselves and the stellar sources
that could be powering them.
We therefore present $N$- and $Q$-band photometry of the Galactic Center,
obtained with the ESO VLT, in order to further investigate the nature of 
the mid-infrared sources.
Additional data at shorter wavelengths is taken from \citet{viehmann2005} to
provide a more complete picture.

Previous studies at mid-infrared wavelengths have been severely limited in
resolution (\citealt{becklin1978} $\sim$2\arcsec, \citealt{gezari1985}
$\ga$1\arcsec, \citealt{stolovy1996} $\sim$0.7\arcsec), however, since the
availability of the mid-infrared camera VISIR at the ESO VLT, the Galactic
Center can be studied in great detail in the  $N$- and $Q$-bands (8-13~$\mu$m
and 17-24~$\mu$m), with a resolution of 0.3\arcsec to 0.6\arcsec (the
diffraction limit of an 8~m telescope), provided that the seeing is good enough.
This corresponds to a linear resolution of 2500 to 5000~AU at the distance of
8~kpc to the Galactic Center.

\section{Observations and data reduction}
\label{obs}

 \subsection{VISIR observations}
 \label{VISOBS}
During the commissioning of the VISIR mid-infrared imager/spectrograph at the
ESO VLT (UT3, Melipal), the Galactic Center was observed on May 7th -- 11th
2004.
Many narrowband filters spanning the whole $N$-band (8-13~$\mu$m), as well as
the 18.7~$\mu$m and 19.5~$\mu$m ($\Delta\lambda=0.88$ and 0.40~$\mu$m) filters
in the $Q$-band were used.
In this paper we concentrate on the 8.6, 11.3, and 12.8~$\mu$m
($\Delta\lambda=0.42$, 0.59, and 0.21~$\mu$m) filters in the $N$-band, since the
highest number of images (and thus the longest integration times) were produced
with these filters.
The images cover a field of view of approximately 32\arcsec$\times$32\arcsec at
a pixel scale of 0.127\arcsec per pixel \citetext{see e.g.\ \citealp{lagage2003}
for further details on VISIR}.
The resulting angular resolution (due to seeing and the diffraction limit)
varies between 0.3\arcsec at 8.6~$\mu$m and approximately 0.6\arcsec in the
$Q$-band.
Each of the images supplied by the VISIR pipeline consists of the average of
(typically) seven chopped (chopper frequency 0.25~Hz, amplitude 12.5\arcsec,
position angle 0\degr) and flat-fielded exposures, with a typical integration
time of 14~s per averaged image.
\footnote{It should be noted that the unfortunate choice of position angle for
chopping results in a negative artifact in the images, since the bright source
IRS~8 lies in the corresponding field of view.}

  \subsubsection{Additional data}
For comparison with shorter wavelengths, $H$-, $K_\mathrm{s}$-, $L$- and
$M$-band (1.6, 2.1, 3.8 and 4.7~$\mu$m) data were taken from
\citet{viehmann2005}.

 \subsection{Data reduction}
\begin{figure}
 \plotone{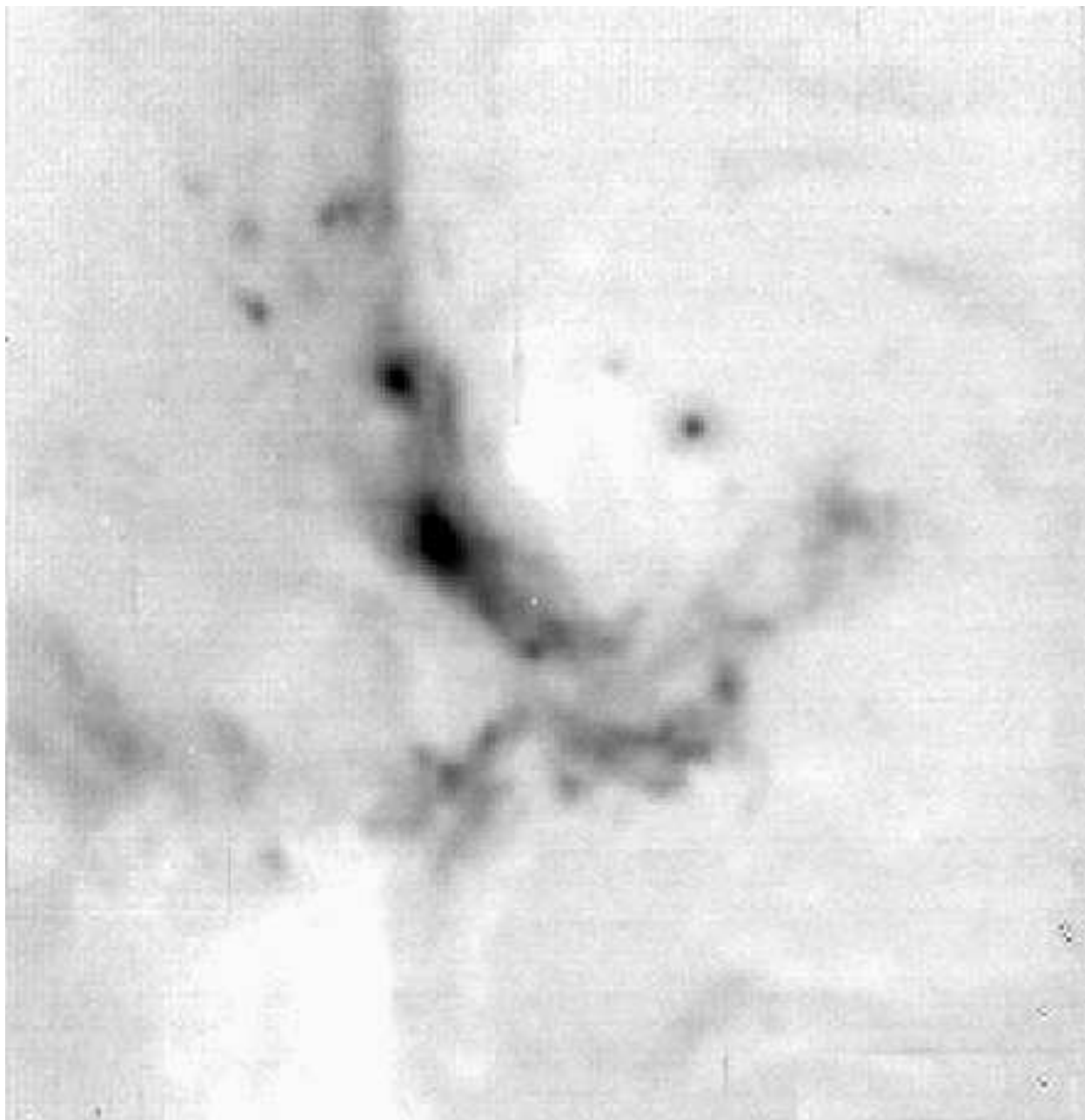}
 \caption{VISIR 19.5~$\mu$m image of the Galactic Center. The field of view is
 32\arcsec$\times$32\arcsec. East is to the left, and north is up.}
 \label{Q3full}
\end{figure}

The chopped and flat-fielded exposures supplied by the pipeline were selected
based on image quality (FWHM and signal-to-noise ratio) and coadded,
resulting in total integration times of 168, 182, 140, 70 and 140 seconds at
8.6, 11.3, 12.8, 18.7 and 19.5~$\mu$m, respectively.
The final 19.5~$\mu$m image is shown in Fig.~\ref{Q3full}, while a three-color
composite of the $N$-band images is shown in Fig.~\ref{rgb}.

  \subsubsection{Photometry}
  \label{calib}
Flux densities for the MIR sources were obtained using aperture photometry with
an aperture of approximately 1\arcsec.
Additional flux densities were obtained with apertures of 2\arcsec and
0.5\arcsec in order to correct for confusion effects for closely separated
sources, and to optimise the background subtraction.
For the purpose of background subtraction, the sky contribution is fitted to an
annulus situated between radii of 1.6\arcsec and 2.1\arcsec (from the center of
the aperture) in the case of the 1\arcsec aperture, and between 2.6\arcsec and
3.1\arcsec or between 1\arcsec and 1.5\arcsec for the 2\arcsec and 0.5\arcsec
apertures, respectively.

Since no suitable calibration stars were observed, the flux density calibration
proved difficult.
Very few $N$- or $Q$-band magnitudes or flux densities of Galactic Center
sources have been published at present.
We used IRS~21 as calibration source, since it is the $N$-band source studied in
greatest detail \citep[e.g.][]{gezari1985, tanner2002}.
We used the flux densities for IRS~21 at 8.8, 12.5, 20.8, and 24.5~$\mu$m
given by \citet{tanner2002}, assuming that the emission from IRS~21 follows the
general shape of the integrated Galactic Center spectrum as measured by ISO
\citep{lutz1996}.
For 8.6~$\mu$m, we also took into account the flux density at 8.7~$\mu$m
published by \citet{stolovy1996}.
Corrections for the different aperture sizes and background subtraction were
applied.
For extinction correction, we assumed A$_\mathrm{V}$=25 \citep{scoville2003,
viehmann2005}, adopting the extinction law of \citet{moneti2001} for the $N$-
and $Q$-bands.

We estimate the total uncertainty (i.e. statistical and systematic) of the
photometry results as $\pm$30\%, rising to $\pm$50\% for the faintest sources
($\lesssim$0.5~Jy).
The dominating uncertainty here is that of the flux density calibration, and
also the correct background subtraction for the faint sources.

\section{Results}
\label{results}

 \subsection{New detections}
 \label{newdet}
\begin{figure}
 \plotone{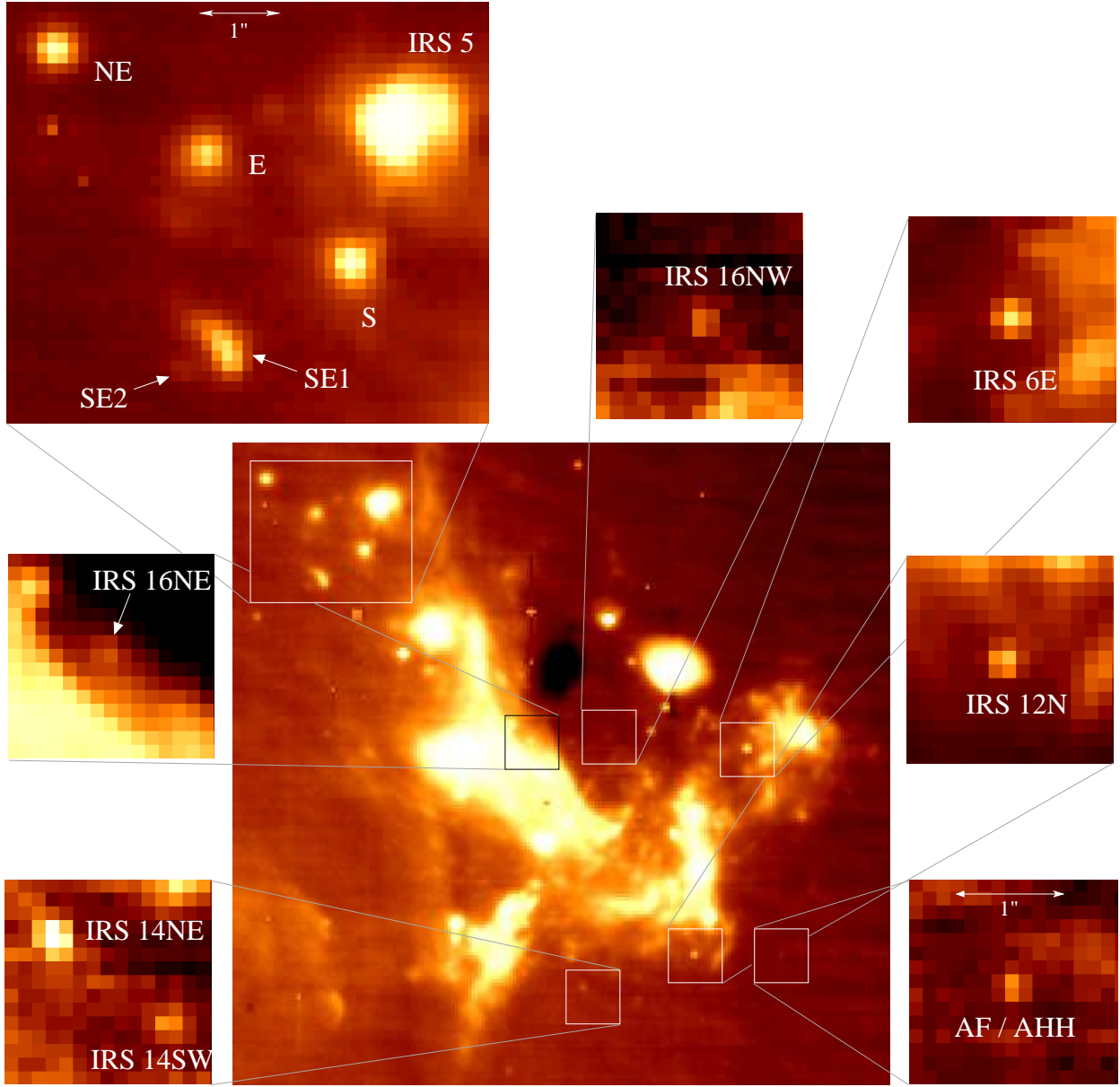}
 \caption{VISIR 8.6~$\mu$m view of the Galactic Center. The field of view is
 approximately 24\arcsec$\times$24\arcsec. The insets show enlarged views of the
 newly detected stars AF/AHH, IRS~6E, IRS~12N, IRS~14NE, IRS~14SW, IRS~16NE and
 IRS~16NW (see section~\ref{newdet}), as well as the intriguing group of bright
 sources located to the east of IRS~5 (upper left, see section~\ref{irs5x}).}
 \label{findthem}
\end{figure}

We have obtained photometry for a total of 35 sources at 19.5~$\mu$m rising to
64 sources at 8.6~$\mu$m.
This is probably the largest number of Galactic Center sources investigated in
the $N$- and $Q$-bands up to now and includes the first detection of the
Galactic Center Helium-emission-line stars IRS~16NE, IRS~16NW and AF/AHH, as
well as the red giants IRS~12N, IRS~14NE and IRS~14SW in the $N$-band, as shown
in Fig.~\ref{findthem}. Additionally, it is now possible to separate the WC9
star IRS~6E \citep{blum1996} from the bright extended emission associated with
IRS~6W. The corresponding data can be found in table~\ref{lotsofnumbers}.

\subsection{SEDs}
\begin{figure}
 \plotone{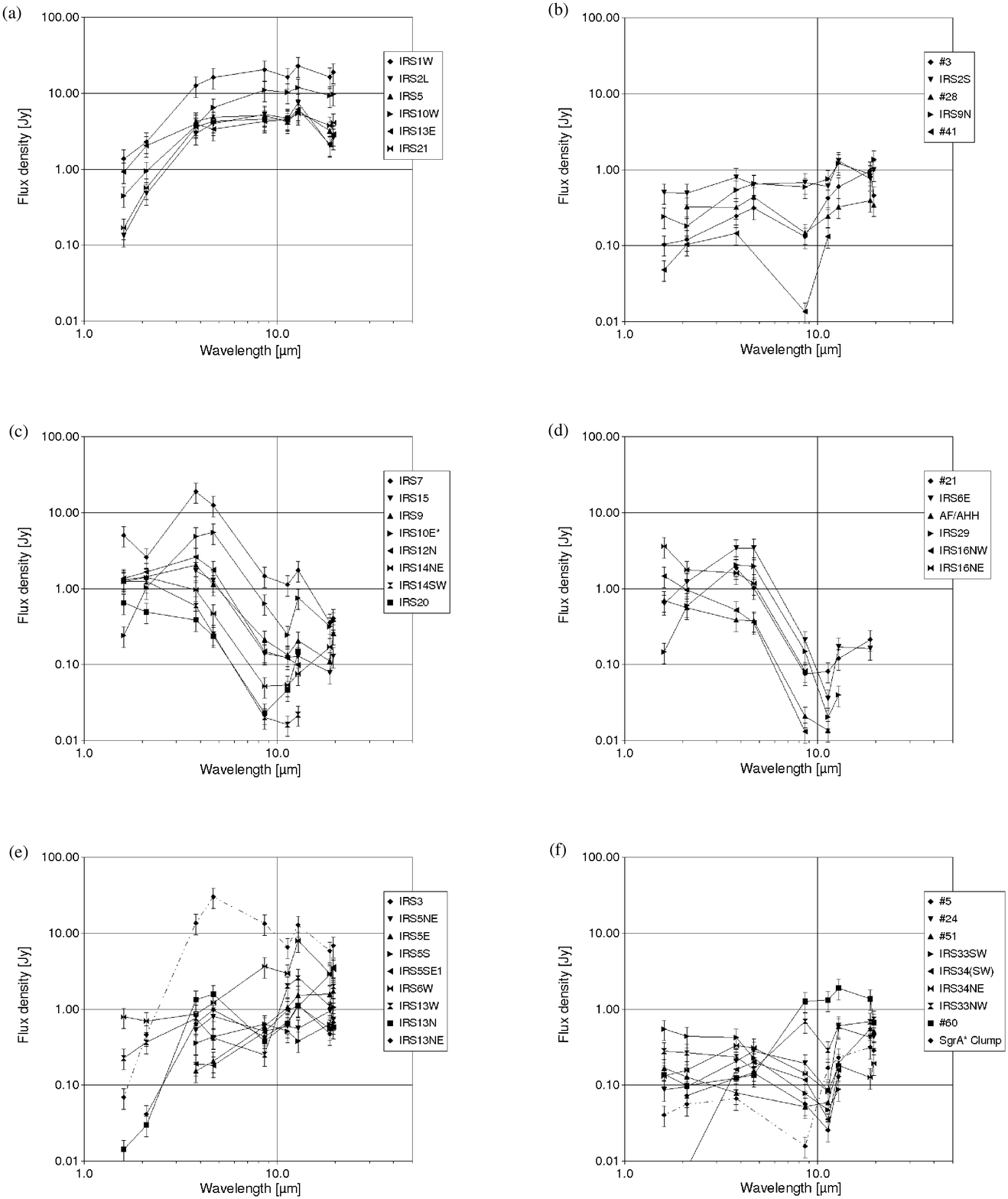}
 \caption{Spectral energy distributions of Galactic Center mid-infrared
 sources. The SEDs have been grouped by type, as described in
 section~\ref{sedtype}: (a) shows the typical luminous bow-shock sources, (b)
 shows lower luminosity bow-shock sources, (c) cool stars, (d) hot stars, while
 (e) and (f) show the SEDs of unclassified sources. In (e) and (f), IRS~3 and
 the Sgr~A* clump are distinguished by a dash-dotted line.}
 \label{SEDs}
\end{figure}

Combining the VISIR data with shorter wavelength data enables us to construct an
unprecedented (for the Galactic Center) number of spectral energy distributions
(SEDs) across the whole 1.6 to 19.5~$\mu$m range. These are shown in
Fig.~\ref{SEDs}.

  \subsubsection{Classification by SED?}
  \label{sedtype}
Preliminary analysis of a few typical cases indicates that these SEDs can
provide interesting information on the source characteristics, since different
source types have SEDs of considerably different shape.
This could possibly allow for preliminary classification of presently
unclassified sources.
The different types of SED found in our sample are as follows:

 \paragraph{I.\ Northern Arm embedded sources\\}
  These are luminous sources that are embedded in the ``Northern Arm'' stream of
  gas and dust, and also in the bar area of the minispiral.
  IRS~1W, 2L, 5, 10W and 21 belong to this group, which shows very red and
  fairly featureless SEDs, with their maximum in the $N$-band, indicating fairly
  low temperatures \citep[approximately 300~K according to][]{gezari1985}.
  These objects are successfully described as bow shock sources
  \citep{tanner2002, tanner2003, tanner2005, rigaut2003, eckart2004,
  geballe2004}.
  IRS~13E, which is situated in the bar area of the minispiral, shows similar
  characteristics, although it is clearly hotter \citep{gezari1996}, with the
  embedded source  identified as a cluster of massive young stars
  \citep[e.g.][]{krabbe1995, najarro1997, eckart2004, maillard2004}.
 \paragraph{II.\ Lower luminosity bow shock sources\\}
  These sources are less luminous and often situated close to the edge of the
  minispiral.
  In almost all cases, they are identified as bow-shock sources by their
  morphology, the bow-shock structure being visible in high-resolution $L$- or
  $M$-band images obtained with adaptive optics \citep[see e.g.][for such
  images]{genzel2003, clenet2004m}.
  A striking example (source 41 in our sample) is presented by
  \citet{clenet2004}.
  These sources show very red SEDs with their maxima longwards of 12~$\mu$m.
  Their characteristics at shorter wavelengths are diverse: some show a fairly
  flat SED between 1.6 and 11.3~$\mu$m, while others exhibit a minimum around
  8.6~$\mu$m.
  These differences are probably due to differences between the type of stars
  powering the bow-shocks (see below), and due to the presence and density of
  the dusty medium that surrounds them and with which they interact.
  We propose that IRS~2S, which has been classified as a cool red giant by
  \citet{blum1996}, belongs to this type of source, since it shows a flat SED
  from 1.6 to 11.3~$\mu$m and brightens considerably at longer wavelengths.
 \paragraph{III.\ Cool stars\\}
  The SEDs of these sources all have their maximum in the 2 to 5~$\mu$m region.
  The SEDs then fall off to the $N$-band, where they appear to flatten out. In
  those cases which are still detectable at wavelengths longer than 8.6~$\mu$m
  (despite decreasing resolution and sensitivity), the $N$- and $Q$-band flux
  densities of this type of source appear to be more or less constant for each
  source.
  This effect is probably due to circumstellar dust emission, which is to be
  expected, since the brightest cool red giants at the Galactic Center have been
  identified as AGB stars \citep{ott1999, clenet2001} and are thus situated in a
  mass-loss phase, creating dust envelopes.
  A \emph{very} weak bow-shock type interaction with the surrounding medium is
  also possible in the case of AGB stars, since they have fairly strong stellar
  winds.
 \paragraph{IV.\ Hot stars\\}
  These sources exhibit SEDs that fall off steeply (steeper than those of cool
  stars) between 5 and 8~$\mu$m.
  Therefore, the $M-N$ color (more particularly $M-m_{8.6\mu\mathrm{m}}$)
  appears to be a criterion for distinguishing hot and cool stars, if
  spectroscopic information is not available.
  The clearly identified hot stars in our sample fall into two distinctive
  subgroups:
  The Galactic Center He-stars IRS~16NE, IRS~16NW and AF/AHH, which are
  frequently classified as LBV or Ofpe/WN9 \citep[e.g.][]{clenet2001}, are
  brightest at short wavelengths with SEDs that fall monotonically towards
  longer wavelengths.
  The dominant source of infrared emission is thus probably photospheric, with a
  wind (or faint bow shock) contribution in the $L$- and $M$-band
  \citep[see][]{viehmann2005}.
  On the other hand, IRS~29 and IRS~6E, which have been classified as WC9
  Wolf-Rayet stars \citep[e.g.][]{blum1996}, exhibit SEDs which peak in the $L$-
  and $M$-bands, while their $H$- and $K$-band flux densities appear to be
  considerably lower.
  The corresponding interpretation is that these stars are surrounded by a
  self-generated dust shell which partly extincts the shorter wavelength
  emission and itself emits strongly at 3-5~$\mu$m
  \citep[see also][]{moultaka2004}.
 \paragraph{V.\ IRS~3\\}
  This source clearly stands apart from all other Galactic Center mid-infrared
  sources.
  While \citet{horrobin2004} classified IRS~3 as a WC5/6 star, \citet{pott2005}
  and R.~Genzel (private communication) have pointed out that this
  identification probably applies to a different, fainter source situated
  approximately 0.12\arcsec to the east of IRS~3.
  This does not rule out that IRS~3 may be a massive Wolf-Rayet star, but
  \citet{pott2005} have also shown that it may be an extreme AGB star.
  The true nature of IRS~3 therefore remains unclear.
  Among the bright $N$-band sources, it is the most compact \citep[see new
  interferometric MIDI VLTI results by][]{pott2005}, apart from IRS~7
  \citep{gezari1985}.
  It is also the brightest source in the $M$-band and one of the brightest over
  the whole mid-infrared region covered here ($L$- to $Q$-band), while it
  appears unremarkable in the $K$-band and very faint at shorter wavelengths.
  The shape of the SED of IRS~3 is thus unlike any of the typical cases
  discussed at present:
  IRS~3 is much fainter than the luminous bow-shock sources in the
  near-infrared, while clearly hotter \citep[approximately 400 to 600~K,
  see][]{gezari1985, gezari1996} and thus showing maximum emission at a shorter
  wavelength ($M$-band vs.\ 10~$\mu$m).
  Compared to luminous cool stars (IRS~7, IRS~10E*), IRS~3 is clearly fainter at
  near-infrared wavelengths and much brighter in the $N$- and $Q$-bands.
  The same applies to the comparison of IRS~3 with the hot stars in our sample.
  However, a certain similarity to the SED of IRS~29 (WC9 type) is discernible:
  The source is faint and obviously extincted in the $H$-band, has a maximum at
  short midinfrared wavelengths ($L$- and $M$-band) and shows a fairly clear
  absorption feature at approximately 10~$\mu$m (see section \ref{silicate}).
  The reason for this similarity is that the emission from both objects comes
  from thick, self-generated dust shells \citep[compare][]{viehmann2005}.

  \subsubsection{Silicate dust}
  \label{silicate}
\begin{figure}
 \plotone{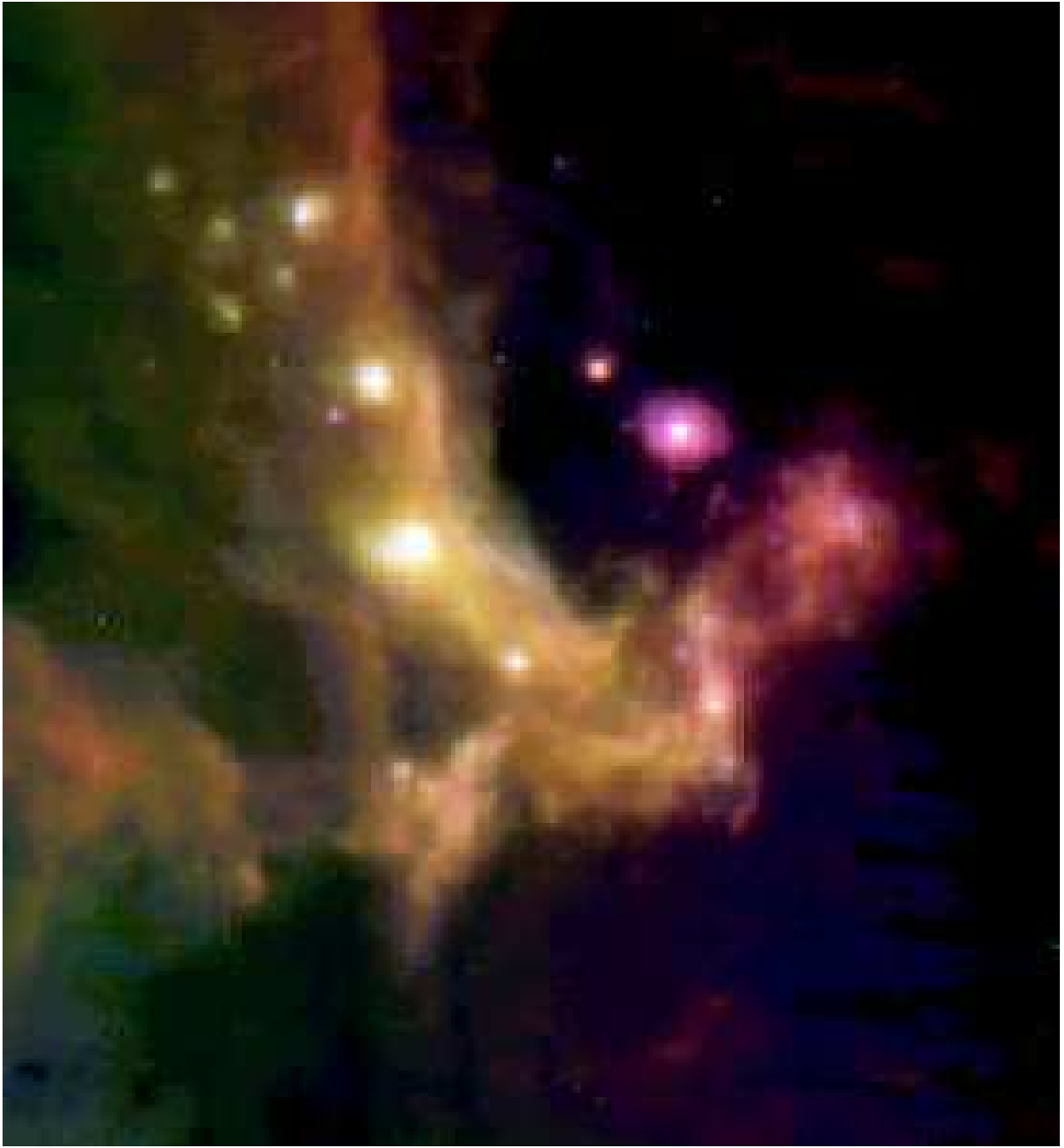}
 \caption{VISIR $N$-band three-color composite view of the Galactic Center. The
 blue channel is 8.6~$\mu$m, green is 11.3~$\mu$m and red is 12.8~$\mu$m.}
 \label{rgb}
\end{figure}

The spectrum between 8.6~$\mu$m and 12~$\mu$m is especially interesting, since
there is a broad interstellar silicate absorption feature at 9.7~$\mu$m
\citep{lutz1996}.
Our data at 11.3~$\mu$m lie on the flank of this absorption; consequently,
sources that are underluminous at this wavelength appear to show additional, and
therefore \emph{intrinsic} silicate absorption, i.e.\ absorption in excess of
that found in the overall ISO spectrum \citep{lutz1996}.
IRS~3 and IRS~6W clearly show this effect, which is also visible in
Fig.~\ref{rgb} (lack of green), as do IRS~34 (both components) and source 24.
The most likely explanation is that these stars are surrounded by (possibly
self-generated) dust shells containing additional silicate dust.
The cool star IRS~10E* and the hot stars IRS~6E and IRS~29, which have been
identified as WC9 Wolf-Rayet stars \citep[e.g.][]{blum1996}, might also show
signs of such absorption, although the steepening of their SED is slight and may
therefore be insignificant.
In the case of the WC9 stars, intrinsic silicate absorption is difficult to
explain, however, one possibility is that the absorption is due to collected
material not originally produced by the stars themselves.
Enhanced, patchy foreground extinction over the whole IRS~3 to IRS~6W area is a
possible alternative explanation.

In contrast, none of the luminous, bow-shock dominated sources shows intrinsic
silicate absorption.
Therefore they appear yellow in the color image shown in Fig.~\ref{rgb}.
It seems that the lower luminosity bow-shock sources also do not show such an
absorption effect.

  \subsubsection{Excess emission at 12.8~$\mu$m}
  \label{high12}
Due to the shape of the Galactic Center spectrum \citep[e.g.][]{lutz1996},
slightly higher flux densities are to be expected at 12.8~$\mu$m.
This is clearly visible in the SEDs obtained from our observations, however,
some sources show a larger excess, which appears to indicate additional
intrinsic emission at 12.8~$\mu$m and must be explained.

While \citet{willner1978} state that the 12.8~$\mu$m \ion{Ne}{2} line emission
at the Galactic Center is extended and not associated with compact mid-infrared
sources, \citet{lacy1979, lacy1980} find that part of the \ion{Ne}{2} emission
comes from compact regions of ionized gas which are indeed associated with
mid-infrared continuum sources. Consequently, \ion{Ne}{2} line emission is a
possible explanation for at least some of these excesses.

In the case of IRS~10E* (and IRS~9 if the small excess of this source is
significant), we attribute the 12.8~$\mu$m excess to a narrow streamer of
enhanced emission running approximately southwards from IRS~5, which probably 
resulted in an insufficient background subtraction (see section~\ref{calib}).
The excess 12.8$\mu$m emission of IRS~2, 6W and 13 could also be the result of
insufficient background subtraction due to the clumpy nature of the extended
emission from this part of the minispiral, which shows quite a complex
structure.

The excess 12.8~$\mu$m emission clearly visible in the SEDs of IRS~3, 7 and
possibly 29 (also visible in Fig.~\ref{rgb}), however, cannot be explained in
this fashion.
Since \citet{lacy1980} state that IRS~3 and IRS~7 do not have associated gas
clouds, it appears unlikely that the excess of these sources is due to
\ion{Ne}{2} line emission.
We therefore suggest that the excess of these sources at 12.8~$\mu$m is due to
continuum emission caused by a faint bowshock-like interaction of the stellar
winds of these stars with the surrounding medium.
Since this is a small effect, the corresponding excess that is to be expected at
11.3~$\mu$m could be masked by intrinsic silicate absorption (see above).
The absence of a clear excess at 18.7~$\mu$m in these sources does not rule out
such an explanation, since the known bow-shock sources (types I. and II. in
section~\ref{sedtype}) show diverse characteristics in the $Q$-band (the
fainter sources have their maxima there, the brighter ones have their maxima in
the $N$-band).

 \subsection{Unusual sources east of IRS~5}
 \label{irs5x}
Among the most intriguing mid-infrared sources at the Galactic Center are four
bright pointlike sources located to the east of the bright Northern Arm source
IRS~5.
The nature of these sources, which we refer to as IRS~5NE, IRS~5E, IRS~5S, and
IRS~5SE (see Fig.~\ref{findthem}), is currently unclear.
They are remarkable in that they are almost as bright as IRS~7 in the $N$-band,
while they appear much less prominent at shorter wavelengths, although they
remain bright in the $L$- and $M$-bands.
Detailed inspection reveals that IRS~5SE, which appears \emph{slightly} extended
at longer wavelengths, is in fact double, consisting of a blue point source to
the east (IRS~5SE2, dominant component in $K$-band images) and a fainter source
to the west, which shows a tail like structure and is the main source of
emission at longer wavelengths (IRS~5SE1, dominant component in $N$- and
$Q$-band).
This may represent an intriguing case of possible interaction, either of the two
sources with each other -- if they lie close together along the line of sight --
or of the western component with the surrounding medium. A further possibility
is that a bow shock generated by the eastern component is superimposed on the
western one, however, their separation makes this unlikely.

All of these sources except IRS~5SE2 show SEDs that are either flat between 3.8
and 20~$\mu$m (IRS~5NE) or increase dramatically towards longer wavelengths.
Therefore, these sources show similar characteristics to identified
lower-luminosity bow-shock sources (e.g. IRS~9N, see section~\ref{sedtype}),
however, their appearance on high-resolution NAOS/CONICA images does not
indicate such a structure, which is normally visible for this type of source
(IRS~5SE1 is the obvious exception here).
It therefore appears that a reliable classification and/or understanding of
these sources can only be achieved with spectroscopic methods or possibly
interferometry using VLTI.

\section{Conclusion}
\label{conc}

We have presented mid-infrared photometry of the Galactic Center from images
obtained with the VISIR camera at the ESO VLT.
In combination with shorter wavelength data \citep{viehmann2005}, we were able
to construct SEDs for an unprecedented number of sources (for the Galactic
Center) in this wavelength regime.
We find that different source types show characteristic differences in the shape
of their SEDs, which should help in the preliminary classification of compact
mid-infrared sources.
In particular, we find that additional, i.e. intrinsic, absorption at
approximately 10~$\mu$m is a sign of dust formation.
Many of the brightest Galactic Center mid-infrared sources, however, appear to
be bow-shock powered and do not show dust formation features in their SEDs.
In addition, we find that hot and cool stars show different
$M-m_{8.6\mu\mathrm{m}}$ colors, with the cool stars showing redder colors, as
is to be expected.

We have reported on four fairly bright $N$-band point sources located to the
east of the bright Northern Arm source IRS~5.
The most likely explanation for their unusual appearance at mid-infrared
wavelengths is that these sources are bow-shock sources of lower luminosity than
the typical Northern Arm sources.

The main uncertainty in our results is due to calibration uncertainties as a
result of the lack of suitable reference observations.
Therefore, mid-infrared observations of the Galactic Center remain necessary in
order to obtain more complete information on the dusty mid-infrared sources in
this intriguing environment.

\acknowledgements

This work was supported in part by the Deutsche Forschungsgemeinschaft
(DFG) via grant SFB 494.


\bibliographystyle{apj}
\bibliography{ms}

\clearpage

\begin{deluxetable}{clrrrrrrrrrrr}
\tabletypesize{\footnotesize}
\tablewidth{0pt}
\tablecaption{Extinction corrected flux densities of compact $N$-band sources.\label{lotsofnumbers}} 
\tablehead{
 \colhead{ID} & \colhead{Name} & \colhead{$\Delta\alpha$(\arcsec)\ \tablenotemark{a}} &
 \colhead{$\Delta\delta$(\arcsec)\ \tablenotemark{a}} & \multicolumn{9}{c}{Flux density (Jy)\ \tablenotemark{b}}\\
 \multicolumn{2}{c}{Wavelength ($\mu$m)} & \nodata & \nodata & 1.6 & 2.1 & 3.8 & 4.7 & 8.6 & 11.3 & 12.8 & 18.7 &
 19.5
}
\startdata
 1 & IRS~1W  & 5.1 & -5.05 & 1.37 & 2.29 & 12.65 & 16.29 & 20.42 & 16.35 & 22.93 & 16.48 & 18.94\\
 2 & \nodata & 5.8 & -5.3 & \nodata & \nodata & \nodata & \nodata & 2.21 & 1.2 & 3.1 & \nodata & \nodata \\
 3\tablenotemark{c} & \nodata & 3.5 & -4.0 & 0.1 & 0.12 & 0.25 & 0.31 & 0.13 & 0.42 & 0.6 & 0.98 & 0.46\\
 4 & \nodata & 3.5 & -4.05 & \nodata & \nodata & \nodata & \nodata & 0.24 & \nodata & \nodata & \nodata & \nodata \\
 5 & \nodata & 5.3 & -6.55 & \nodata & 0.07 & 0.12 & 0.15 & 0.06 & 0.03 & 0.13 & \nodata & \nodata \\
 6 & IRS~2L  & -3.6 & -9.45 & 0.13 & 0.48 & 2.98 & 3.98 & 5.26 & 4.72 & 7.49 & 2.05 & 2.58\\
 7 & IRS~2S \tablenotemark{c} & -4.0 & -11.05 & 0.5 & 0.49 & 0.8 & 0.64 & 0.68 & 0.61 & 1.3 & 0.75 & 1.01\\
 8 & IRS~3  & -2.45 & -1.8 & 0.07 & 0.46 & 13.61 & 30.2 & 13.42 & 6.57 & 12.85 & 5.85 & 6.88\\
 9 & \nodata & -2.25 & -3.3 & \nodata & \nodata & \nodata & \nodata & 0.21 & 0.03 & 0.31 & \nodata & \nodata \\
10 & \nodata & -1.0 & -1.65 & \nodata & \nodata & \nodata & \nodata & 0.14 & 0.06 & 0.14 & 0.23 & \nodata \\
11 & IRS~4  & 10.3 & -11.5 & \nodata & \nodata & \nodata & \nodata & 0.11 & 0.19 & 0.45 & 0.48 & 0.61\\
12 & \nodata & 9.8 & -14.4 &	\nodata & \nodata & \nodata & \nodata & 0.14 & 0.27 & 0.63 & 0.77 & 0.76\\
13 & IRS~5  & 8.5 & 4.15 & \nodata &	\nodata & 4.27 & 4.88 & 5.11 & 4.21 & 5.85 & 3.21 & 2.85\\
14 & IRS~5NE  & 12.7 & 5.0 & \nodata & \nodata & 0.53 & 0.8 & 0.51 & 0.6 & 0.56 & 0.95 & 0.74\\
15 & IRS~5E  & 10.8 & 3.7 & \nodata & \nodata & 0.15 & 0.21 & 0.59 & 1.06 & 1.54 & 1.58 & 1.72\\
16 & \nodata & 10.0 & 4.25 & \nodata & \nodata & \nodata & \nodata & 0.04 & \nodata & \nodata & \nodata & \nodata \\
17 & \nodata & 11.15 & 2.9 & \nodata & \nodata & \nodata & \nodata & 0.05 & \nodata & \nodata & \nodata & \nodata \\
18 & IRS~5S  & 9.0 & 2.35 & \nodata & \nodata & 0.36 & 0.43 & 0.63 & 0.51 & 0.38 & 0.63 & 1.06\\
19 & IRS~5SE1  & 10.55 & 1.3 & \nodata & \nodata & 0.19 & 0.18 & 0.5 & 0.97 & 1.1 & 2.84 & 3.58\\
20 & IRS~5SE2  & 10.95 & 1.1 & \nodata & \nodata & \nodata & \nodata & 0.04 & 0.09 & 0.13 & \nodata & \nodata \\
21 & \nodata & 12.8 & -0.15 &	\nodata & \nodata & 1.88 & 1.03 & 0.08 & 0.08 & 0.12 & 0.21 & \nodata \\
22 & IRS~6W  & -7.5 & -4.25 & 0.79 & 0.7 & 0.87 & 1.21 & 3.67 & 2.95 & 8.0 & 2.95 & 3.37\\
23 & IRS~6E  & -5.3 & -4.8 & 0.63 & 1.22 & 3.42 & 3.44 & 0.21 & 0.04 & 0.17 & 0.16 & \nodata \\
24 & \nodata & -7.35 & -7.25 & 0.09 & 0.1 & 0.21 & 0.28 & 0.19 & 0.08 & 0.58 & 0.43 & 0.43\\

25 & IRS~7  & 0.0 & 0.0 & 5.04 & 2.58 & 18.97 & 12.59 & 1.47 & 1.14 & 1.75 & 0.36 & 0.41\\

26 & \nodata & -0.25 & 2.3 & \nodata & \nodata & \nodata & \nodata & 0.05 & 0.03 & 0.11 & 0.07 & 0.11\\
27 & IRS~15  & 1.1 & 5.65 & \nodata & \nodata & 1.71 & 1.28 & 0.14 & 0.12 & 0.13 & 0.08 & 0.13\\
28\tablenotemark{c} & \nodata & 3.65 & 1.55 & \nodata & 0.33 & 0.32 & 0.44 & 0.15 & 0.24 & 0.32 & 0.4 & 0.34\\
29 & IRS~9  & 5.4 & -11.95 & 1.31 & 1.43 & 2.06 & 1.15 & 0.21 & 0.13 & 0.21 & 0.11 & 0.26\\
30 & IRS~9N \tablenotemark{c} & 5.4 & -11.45 & 0.24 & 0.18 & 0.54 & 0.65 & 0.59 & 0.76 & 1.21 & 0.87 & 1.37\\
31 & \nodata & 3.5 & -10.9 & 0.01 & \nodata & 0.24 & \nodata & 3.68 & 2.6 & 3.58 & 1.4 & 2.94\\
32 & \nodata & 4.05 & -12.9 & 0.02 & 0.02 & 0.27 & 0.45 & 2.93 & 2.77 & 3.9 & 2.13 & 5.51\\
33 & \nodata & 5.0 & -16.0 & 0.07 & 0.06 & 0.07 & \nodata & 0.82 & 0.94 & \nodata & \nodata & \nodata \\
34 & \nodata & 3.4 & -16.9 & \nodata & \nodata & \nodata & \nodata & 0.03 & \nodata & \nodata & \nodata & \nodata \\
35 & IRS~10W  & 6.35 & -0.5 & 0.45 & 0.95 & 3.72 & 6.49 & 11.06 & 10.31 & 11.85 & 9.32 & 9.75\\
36 & IRS~10E*  & 7.5 & -1.45 & 0.24 & 1.04 & 4.9 & 5.5 & 0.64 & 0.25 & 0.75 & 0.32 & 0.37\\
37 & \nodata & 6.55 & 2.05 & \nodata & \nodata & \nodata & \nodata & 0.1 & \nodata & \nodata & \nodata & \nodata \\
38 & \nodata & 7.8 & -0.4 & 0.05 & 0.04 & 0.05 & \nodata & 0.21 & 0.38 & \nodata & \nodata & \nodata \\
39 & \nodata & 6.2 & -8.45 & \nodata & \nodata & \nodata & \nodata & 0.46 & 0.56 & 1.67 & 0.99 & 0.71\\
40 & \nodata & 6.1 & -9.55 & \nodata & \nodata & \nodata & \nodata & 0.48 & 0.12 & 0.76 & \nodata & \nodata \\
41\tablenotemark{c} & \nodata & 3.2 & -2.2 & 0.05 & 0.1 & 0.15 & \nodata & 0.01 & 0.13 & \nodata & \nodata & \nodata \\
42 & IRS~12N  & -3.5 & -12.45 & 1.37 & 1.66 & 2.62 & 1.77 & 0.15 & 0.12 & 0.1 & \nodata & \nodata \\
43 & \nodata & -3.8 & -14.9 & \nodata & \nodata & \nodata & \nodata & 0.02 & 0.03 & \nodata & \nodata & \nodata \\
44 & IRS~13E  & -3.35 & -7.1 & 0.92 & 2.05 & 3.96 & 3.37 & 4.28 & 4.49 & 6.16 & 2.11 & 2.89\\
45 & IRS~13W  & -4.2 & -7.6 & 0.23 & 0.37 & 0.76 & 0.42 & 0.25 & 2.02 & 2.59 & 1.1 & 2.0\\
46\tablenotemark{c} & \nodata & -2.65 & -8.0 & \nodata & \nodata & \nodata & \nodata & 0.36 & 0.24 & 0.74 & \nodata & \nodata \\
47 & IRS~13N  & -3.15 & -6.5 & 0.01 & 0.03 & 1.34 & 1.58 & 0.38 & 0.65 & 1.11 & 0.55 & 0.58\\
48 & IRS~13NE  & -2.85 & -7.0 & \nodata & 0.04 & 0.64 & 0.98 & 0.43 & 0.67 & 1.13 & 0.47 & 0.67\\
49 & IRS~14NE  & 0.7 & -13.75 & 1.24 & 1.41 & 0.97 & 0.47 & 0.05 & 0.05 & 0.08 & 0.17 & \nodata \\
50 & IRS~14SW  & -0.4 & -14.5 & 1.25 & 1.25 & 0.6 & 0.25 & 0.02 & 0.02 & 0.02 & \nodata & \nodata \\
51 & \nodata & 1.1 & -12.65 & 0.17 & 0.13 & 0.08 & \nodata & 0.05 & 0.06 & 0.2 & 0.55 & 0.52\\
52 & AF/AHH  & -6.65 & -12.45 & 0.7 & 0.56 & 0.39 & 0.38 & 0.02 & 0.01 & \nodata & \nodata & \nodata \\
53 & IRS~20  & -2.2 & -11.95 & 0.65 & 0.49 & 0.39 & 0.24 & 0.02 & 0.05 & 0.15 & \nodata & \nodata \\
54 & IRS~21  & 2.15 & -8.35 & 0.17 & 0.57 & 3.58 & 4.29 & 4.56 & 4.57 & 5.44 & 3.76 & 4.1\\
55 & IRS~29  & -1.75 & -4.2 & 0.15 & 0.6 & 2.04 & 1.98 & 0.15 & 0.02 & 0.04 & \nodata & \nodata \\
56 & IRS~33SW  & -0.4 & -8.7 & 0.55 & 0.44 & 0.42 & 0.23 & 0.08 & 0.05 & 0.09 & \nodata & \nodata \\
57 & IRS~34(SW)  & -4.05 & -4.0 & \nodata & 0.01 & 0.16 & 0.2 & 0.12 & 0.04 & 0.16 & \nodata & \nodata \\
58 & IRS~34NE  & -3.65 & -3.55 & 0.13 & 0.16 & 0.33 & 0.31 & 0.14 & 0.08 & 0.18 & 0.13 & 0.19\\
59 & IRS~33NW  & -0.9 & -7.25 & 0.28 & 0.27 & 0.23 & 0.16 & 0.69 & 0.29 & 0.61 & 0.68 & 0.73\\
60 & \nodata & -1.65 & -6.15 & 0.14 & 0.1 & 0.12 & 0.14 & 1.27 & 1.32 & 1.9 & 1.37 & 0.66\\
61 & IRS~16NW  & -0.1 & -4.45 & 1.47 & 0.96 & 0.52 & 0.36 & 0.01 & \nodata & \nodata & \nodata & \nodata \\
62 & IRS~16NE  & 2.75 & -4.7 & 3.61 & 1.77 & 1.61 & 1.18 & 0.08 & \nodata & \nodata & \nodata & \nodata \\
63 & \nodata & 1.25 & -4.65 & \nodata & \nodata & \nodata & \nodata & 0.01 & 0.01 & 0.05 & \nodata & \nodata \\
64 & SgrA* Clump  & -0.25 & -5.65 & 0.04 & 0.06 & 0.07 & \nodata & 0.02 & 0.17 & 0.23 & 0.32 & 0.29\\
\enddata
\tablenotetext{a}{\ Position offsets are relative to IRS~7 (17$^\mathrm{h}$45$^\mathrm{m}$40\fs052,
 -29\arcdeg00\arcmin33\farcs659, see \citealt{reid1999,reid2003}) and are accurate to $\pm$0.15\arcsec.}
\tablenotetext{b}{\ Photometry is accurate to $\pm$30\%, or $\pm$50\% for sources fainter than $\sim$0.5~Jy. The
 corresponding aperture size is approximately 1\arcsec.}
\tablenotetext{c}{\ Lower luminosity bow-shock source, see section~\ref{sedtype}.}
\end{deluxetable}

\clearpage

\end{document}